\documentclass[a4paper,fleqn]{cas-sc}

\usepackage[numbers,sort&compress]{natbib}
\usepackage{bm}
\usepackage{tikz}
\usepackage{subcaption}
\usetikzlibrary{decorations.pathreplacing}
\usetikzlibrary{patterns}

\newcommand{\re}{\text{Re}}
\newcommand{\im}{\text{Im}}

\newcommand{\rp}{\textbf{R}_p}
\newcommand{\qp}{\textbf{Q}_p}
\newcommand{\kf}{\textbf{K}_f}

\newcommand{\uok}{\cup_\text{OK}}

\begin{document}
\let\WriteBookmarks\relax
\def\floatpagepagefraction{1}
\def\textpagefraction{.001}
\shorttitle{Amplitude by Ptychography}
\shortauthors{Ducharme and Clark}

\title [mode = title]{Full complex wavefunction reconstruction with direct-interference ptychography}                      
\tnotemark[1]

\tnotetext[1]{This research is funded by a Royal Society University Research Fellowship (URF\textbackslash R1 \textbackslash 221270) and additional Royal Society funding (RF \textbackslash ERE \textbackslash 221035).}


\author[1]{A. Ducharme}[orcid=0000-0003-2765-1455]
\cormark[1]
\ead{andrew.ducharme@york.ac.uk}

\credit{Conceptualization, analysis, simulations, original manuscript preparation}

\affiliation[1]{organization={School of Physics, Engineering and Technology},
                addressline={University of York}, 
                city={York},
                postcode={YO10 5DD}, 
                country={United Kingdom}}

\author[1]{L. Clark}[orcid=0000-0001-8245-162X]

\credit{Validation, manuscript revision, funding acquisition}

\cortext[cor1]{Corresponding author}


\begin{abstract} 
We demonstrate the separable reconstruction of the sample wavefunction amplitude and phase using direct-interference ptychography in scanning transmission electron microscopy. We reconstruct the sample wavefunction in simulated and experimental data under the Weak Object Approximation (WOA) with an extension of single-sideband (SSB) ptychography using information in the double- and triple-overlap regions. The contrast transfer function for the amplitude information is equivalent to the ideal contrast transfer function of Center of Mass imaging (COM). Our results support reporting the imaginary part of SSB reconstructions.
\end{abstract}




\maketitle

\section{Introduction}
Computational phase retrieval has long been a key tool for maximizing the analytical potential of a scanning transmission electron microscope (STEM) \cite{rodenburg_ptychography_2008,ophus_four-dimensional_2019, orchowskiElectronHolographySurmounts1995a, okeefe_sub-angstrom_2001}. Electron detectors can only record the incident current or count the number of incident electrons, measuring only the amplitude of incoming wavefront. The amplitude signal scales as approximately $Z^{1.6-2}$ \cite{hartel_conditions_1996}, where $Z$ is the atomic number of the elements comprising the sample under study. In a typical STEM experimental geometry, microscopists cannot observe the more sensitive phase signal that scales as approximately $Z^{0.6-0.7}$ \cite{kirkland_advanced_2020, linck_electron_2006}. Computational phase retrieval therefore permits the identification of weakly-scattering low-atomic-number species \cite{yang_simultaneous_2016} particularly key in dose-limited samples prevalent in the life sciences without significant STEM customization \cite{yu_dose-efficient_2025,ilett_analysis_2020, pei_cryogenic_2023}.

Iterative electron ptychography is a very successful computational phase retrieval technique for materials science problems where the sample is beam-insensitive \cite{jiang_electron_2018,chen_electron_2021, clark_electron_2025}. At high electron dose, iterative ptychography is capable of exactly reconstructing the sample wavefunction amplitude and phase, deconvolving any probe contributions from the object reconstruction. However, iterative ptychography exhibits worsening information transfer at lower doses necessary to image beam-sensitive materials \cite{ma_information_2024,varnavides_beyond_2026}. 
The iterative reconstruction algorithm can be trapped in a local minimum and produce a less-faithful reconstruction of the sample compared to non-iterative techniques at the same dose \cite{kucukoglu_low-dose_2024, chennit_investigating_2025, dearg_stability_2025}. Thus, it is not surprising to see a concomitant resurgence of interest in non-iterative computational phase retrieval \cite{bennemann_detective_2026,ma_information_2026,ma_using_2026,varnavides_beyond_2026} as STEM pushes towards lower-dose applications \cite{ilett_analysis_2020}. 

\begin{figure}
	\centering
	\includegraphics[width=\textwidth]{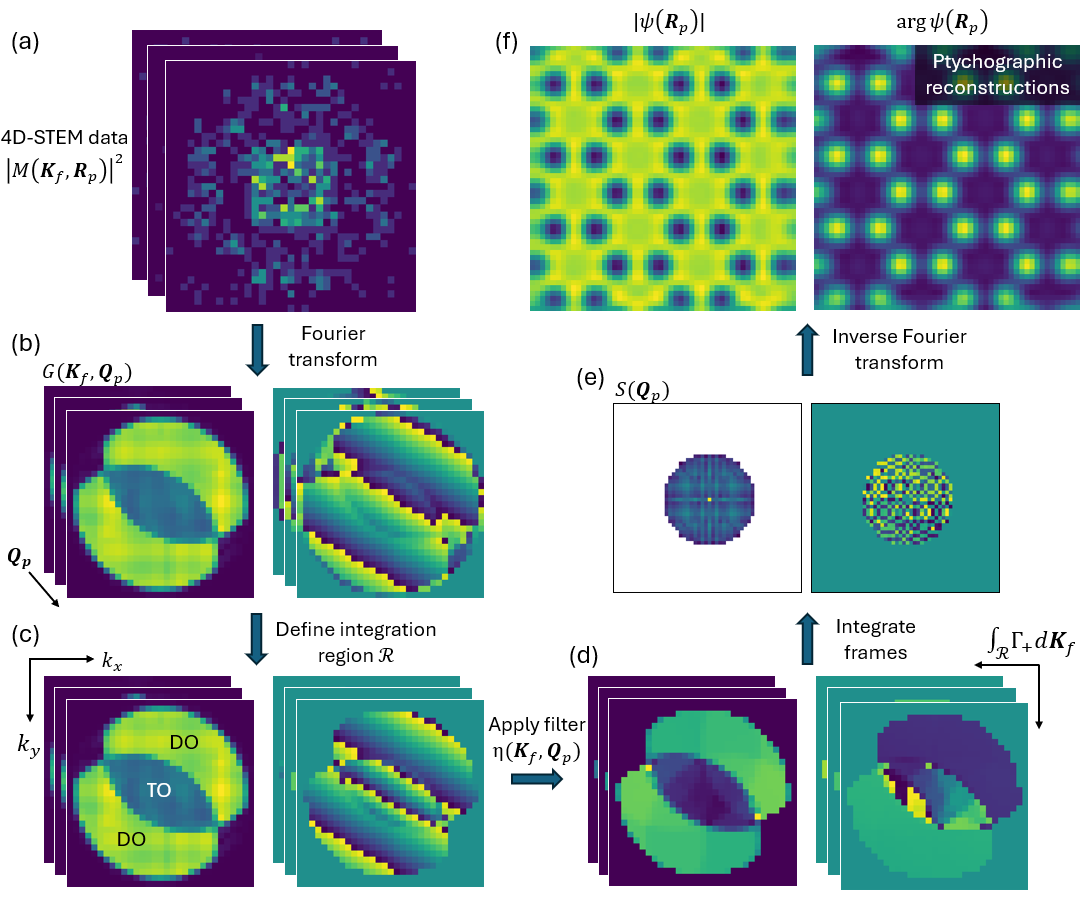}
	\caption{A schematic of the direct-interference ptychography reconstruction algorithm. (a) 4D-STEM data is Fourier transformed with respect to the probe position, creating (b) a dataset in doubly-reciprocal space. After (c) isolating the data containing weak sample information, (d) aberration corrections are applied via a filter $\eta$ before integrating to reduce the data to (e) a two-dimensional complex-valued array. This array is inverse Fourier transformed to produce (f) reconstructions of the sample amplitude and phase.}
	\label{fig:overview}
\end{figure}

Single-sideband ptychography (SSB) is the oldest of a group of closely-related non-iterative computational phase retrieval techniques that are inherently well-suited for low-dose measurements. The 4D-STEM dataset has dimensions $(\kf,\rp) = (k_x,k_y,x,y)$, where $\kf$ and $\rp$ are the coordinates at the detector plane and of the probe position respectively. Let $\qp = (q_x,q_y)$ be the reciprocal coordinate to $\rp$. Under the weak phase object approximation (WPOA), all phase contrast is isolated to a region in $(\kf, \qp)$-space defined by the shape of the probe-forming aperture \cite{rose_nonstandard_1976, oleary_contrast_2021,ma_information_2026}. This domain, depicted in the lower-left corner of Fig. \ref{fig:overview} (and in more detail in Fig. \ref{fig:clark_zone}), is the union of the double-overlap and triple-overlap regions. SSB \cite{rodenburg_experimental_1993,pennycookEfficientPhaseContrast2015}, ptychographic matched illumination detector interferometry STEM (PMIDI-STEM) \cite{yang_enhanced_2016} (also called phase-compensated SSB \cite{varnavides_abcs_2026} or simply direct ptychography \cite{you_gapfree_2026}), optimal bright field STEM (OBF) \cite{ooe_ultra-high_2021}, and sideband-masked center-of-mass (SBm-COM) \cite{ding_defocus_2025} only use a subset of this region to generate a phase reconstruction. This efficiently ignores parts of the 4D-STEM dataset that do not contain phase information, \textit{i.e.}, for the purpose of phase retrieval, only contain noise. These techniques are known for their stability with increasing sample phase in spite of their basis in the WPOA \cite{plamann_electron_1998, clark_effect_2023}, with interest lying in their use for live visualization of sample structure and low-dose measurements \cite{lalandec_robert_benchmarking_2025,li_improving_2025,dearg_stability_2025, varnavides_abcs_2026,bennemann_detective_2026,ooe_dose-efficient_2025}.

To separate these closely-related techniques from other non-iterative phase reconstruction methods, we define direct-interference ptychography (DIP) as computational phase retrieval using only data from a strict subset of the double- and triple-overlap regions. In the WPOA, such techniques yield contrast solely from the interference of the aperture disks in $(\kf,\qp)$-space. A schematic of their shared reconstruction algorithm is provided in Fig. \ref{fig:overview}. Implementations of DIP differ in the portions of DO and TO regions $\mathcal{R}$ used and how contrast is modulated and aberrations are corrected through postprocessing filters $\eta(\kf,\qp)$. Other common phase contrast techniques use different portions of the 4D-STEM dataset: Wigner distribution deconvolution (WDD) ptychography \cite{rodenburg_theory_1992} also uses slices of 4D-STEM data along $\kf$ in $(\kf,\qp)$-space, but does not restrict itself to the DO and TO regions so is not a DIP technique. Neither are tilt-corrected Bright Field (tcBF)/parallax \cite{nguyen_4d-stem_2016,yu_dose-efficient_2025} and aberration-corrected Bright Field (acBF) \cite{ma_information_2026} imaging, which use slices along $\qp$ in $(\kf, \qp)$-space, or differential phase contrast (DPC) \cite{dekkers_differential_1974}, integrated center of mass (iCOM), and their variants \cite{lazic_phase_2016}, which use slices along $\kf$ in $(\kf, \rp)$-space. 

In this manuscript, we demonstrate the non-iterative direct-interference ptychographic retrieval of both the sample wavefunction amplitude and phase under a weak object approximation (WOA). We show that the desired sample wavefunction amplitude and phase can be separably reconstructed. The sample amplitude can be reconstructed independently of the sample phase regardless of the probe aberrations. The sample phase can be reconstructed independently of the sample amplitude when the measurement is performed in focus. This result replicates the separable measurement of amplitude and differential phase contrast by symmetric and anti-symmetric segmented detectors using a symmetric probe under the WOA in DPC \cite{seki_linear_2022} and complements quantitative wavefunction retrieval by iterative multislice electron ptychography \cite{denzer_quantitative_2026, denzer_quantitative_2026-1}. We derive the analytical forms of our phase and amplitude reconstructions in a Fourier optics model of the STEM, finding the amplitude reconstruction is the convolution of the sample wavefunction amplitude with the intensity of the STEM probe. This implies ideal amplitude contrast transfer for a diffraction-limited linear imaging system with a circular aperture \cite{born_principles_2019} (distinct from the unitary transfer of iterative ptychography at infinite dose).

\begin{figure}
	\centering
	\includegraphics[width=\textwidth]{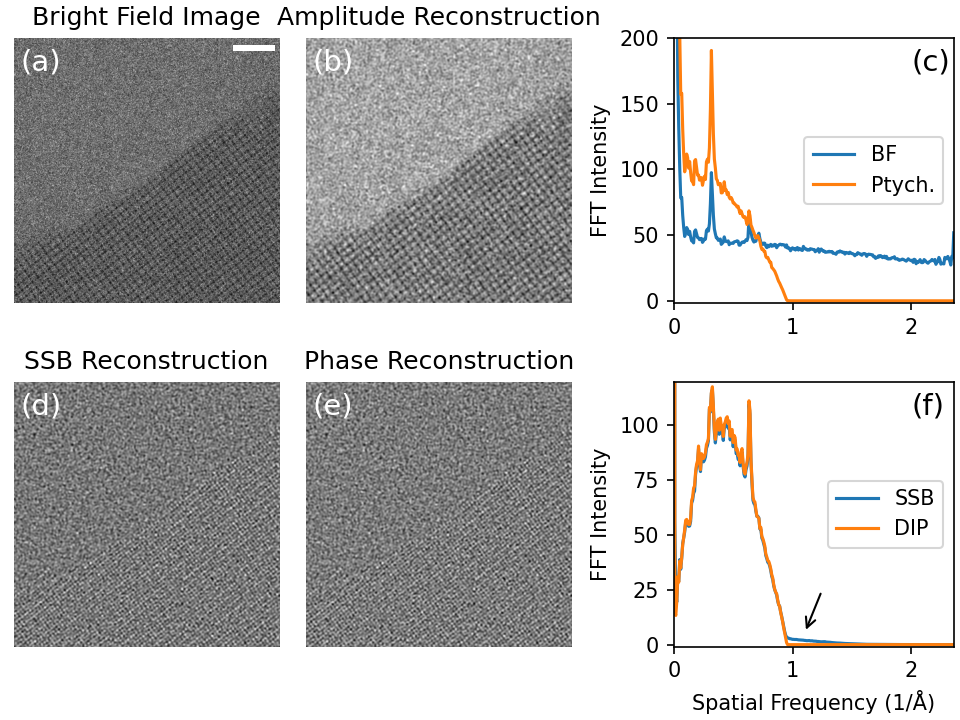}
	\caption{Reconstruction of lead halide perovskite $\text{MAPbI}_3$ data from Ref. \cite{yuan_atomically_2025}. This data was reported to have an electron dose of 33 $\text{e}^-/\text{\AA}^2$. Subfigures (a-c) illustrate amplitude imaging: (a) is the BF STEM image acquired by radial integration up to the semi-convergence angle. The scale bar is 3 nm. (b) is the wavefunction amplitude reconstruction by our DIP method. (c) compares the radially-integrated discrete Fourier transforms of the amplitude images in (a) and (b). Subfigures (d-f) illustrate phase reconstruction: (d) is the conventional SSB reconstruction, (e) is the wavefunction phase reconstruction by our DIP method, and (f) is the comparison of the radially-integrated discrete Fourier transforms of both phase reconstructions. An arrow guides the eye to unexpected signal beyond the $2\alpha$ in the SSB reconstruction.}
	\label{fig:prelim_example}
\end{figure}

The noise filtering inherent to DIP techniques form more interpretable amplitude-contrast images, such as in Fig. \ref{fig:prelim_example}, an analysis of an ultra-low-dose (33 $\text{e}^-/\text{\AA}^2$) measurement of the lead halide perovskite $\text{MAPbI}_3$ performed by Yuan \textit{et al.} \cite{yuan_atomically_2025} and accessible in Ref. \cite{yuan_data_2025}. Our amplitude reconstruction has a much clearer distinction between sample and vacuum relative to the conventional Bright Field (BF) STEM image generated by incoherently summing all counts up to the semi-convergence angle at the detector. The discrete Fourier transforms of the amplitude images plotted in Fig. \ref{fig:prelim_example}(c) demonstrate the concentration of signal within the $2\alpha$ information limit. The phase reconstructions are far more similar, an observation supported by their Fourier transforms in Fig. \ref{fig:prelim_example}(f). Surprisingly, the conventional SSB reconstruction has a non-zero Fourier transform beyond its information limit at $2\alpha$. Our reconstruction properly loses contrast transfer beyond $2\alpha$, providing a minor benefit in contrast transfer. We find this is due to a SSB convention of reporting the phase of the complex-valued reconstruction instead of its imaginary part.

This work is structured as follows. In Sections \ref{classic_theory} and \ref{recon_theory}, we develop the linear Fourier optical theory of our method. In Section \ref{classic_theory}, we derive the conventional SSB ptychography theory under the WOA. In Section \ref{recon_theory}, we analytically compute the form of the direct-interference ptychography reconstruction. This generates as a byproduct the contrast transfer and point spread functions for our technique (see Section \ref{ctfs_and_ptfs}). In Section \ref{discussion}, we present reconstructions of simulated measurements of $\text{SrTiO}_3$ data and discuss their performance relative to BF-STEM and SSB ptychography before concluding.

\section{STEM in the Weak Object Approximation} \label{classic_theory}

In this and the next section, we derive a closed-form equation for a direct-interference ptychography reconstruction in terms of the sample wavefunction and the STEM probe intensity under the weak object approximation. This section models the propagation of the electron beam to the detector and the first postprocessing step, a Fourier transform with respect to the probe position. We choose this division to match the typical presentations of SSB theory in the WPOA and to emphasize the slight changes resulting from using the WOA \cite{rodenburg_experimental_1993,pennycookEfficientPhaseContrast2015}. A similar derivation to our own can be found in Ref. \cite{seki_linear_2022}.

We model the STEM experimental geometry as a complex-valued electron probe $a(\textbf{R})$ in real space $\textbf{R}=(x,y)$ illuminating a sample with complex-valued wavefunction $\psi(\textbf{R})$. The illumination is of a wavelength $\lambda$ and semi-convergence angle $\alpha$. This probe is rastered across the sample to various positions $\textbf{R}_p$. Under Fourier optics, at each probe position, there exists a far-field diffraction plane $\textbf{K}_f = (k_x,k_y)$ where the probe electron wavefunction can be described by 
\begin{equation}
    M(\textbf{K}_f, \textbf{R}_p) = FT[a(\textbf{R} - \textbf{R}_p)\psi(\textbf{R}), \textbf{R}] = [A(\textbf{K}_f) e^{2\pi i \textbf{K}_f \cdot \textbf{R}_p}] \otimes_{\textbf{K}_f} \Psi(\textbf{K}_f)
\end{equation}
after employing the Fourier shift and convolution theorems. Uppercase letters are used to denote the Fourier transforms $FT$ of real-space functions denoted by lowercase letters. This formalism uses the Fourier transform convention 
\begin{align}
    F(\bm k) &= FT[f(\bm x), \bm x] = \int f(\bm x) e^{2\pi i \bm k \cdot \bm x} d \bm x \\
    f(\bm x) &= FT^{-1}[F(\bm k), \bm x] = \int F(\bm k) e^{-2\pi i \bm k \cdot \bm x} d \bm k.
\end{align}
$\otimes_{\kf}$ denotes a convolution with respect to the variable $\kf$. In typical STEM operation, the probe-forming aperture $A(\textbf{K}_f)=|A(\textbf{K}_f)|e^{i\chi (\kf)}$ is approximately circular, with amplitude
\begin{equation}
    |A(\textbf{K}_f)| = \begin{cases}
    1, |\textbf{K}_f| < \frac{\alpha}{\lambda} \\
    0, |\textbf{K}_f| > \frac{\alpha}{\lambda}
    \end{cases}.
\end{equation}
Geometric aberrations due to misalignments of the microscope describe the phase structure of $A(\kf)$
\begin{equation}
    \chi(\kf) = \chi(k, \phi) = \frac{2\pi}{\lambda} \sum_{n=1}^\infty \sum_{m=0}^{n+1} \frac{1}{n+1} C_{n,m} (k\lambda)^{n+1} \cos[m(\phi-\phi_{n,m})]
\end{equation}
where $k=\sqrt{k_x^2 + k_y^2}$, $\phi = \arctan \frac{k_y}{k_x}$, and $C_{n,m}, \phi_{n,m}$ are the aberration coefficients and rotation angles in Krivanek notation \cite{krivanek_towards_1999}.

The 4D-STEM detector records the intensity of $M$
\begin{equation} \label{detector_intensity}
\begin{aligned}
    |M(\textbf{K}_f, \textbf{R}_p)|^2 &= M^*(\textbf{K}_f, \textbf{R}_p) M(\textbf{K}_f, \textbf{R}_p) \\
    &= \iint  A(\textbf{K'}) \Psi(\textbf{K}_f - \textbf{K'}) A^* (\textbf{K''}) \Psi^* (\textbf{K}_f - \textbf{K''}) \exp[2\pi i \textbf{R}_p \cdot (\textbf{K'} - \textbf{K''})] d\textbf{K'}d\textbf{K''},
\end{aligned}
\end{equation}
which, without any additional approximations made yet, describes all diffraction, therefore is generally intractable. To better understand measured diffraction patterns, we desire to transform this equation such that the sample information is no longer being integrated. Conventionally, one would use the weak phase object approximation (WPOA). We instead proceed using the weak object approximation (WOA)
\begin{equation}
    \psi(\bm R) = e^{f(\bm R)} \approx 1 + \psi_a(\textbf{R}) + i\psi_\phi(\textbf{R}),
\end{equation}
namely that the wavefunction amplitude and phase are sufficiently weak such that the exponential is approximately equal to its first-order Taylor expansion. The WPOA further assumes $\psi_a(\bm R) = 0$. The real and imaginary parts of $f(\bm R)$
\begin{equation}
    f(\bm R) = \psi_a(\bm R) + i \psi_\phi (\bm R)
\end{equation}
are so defined because they are directly related to the amplitude and phase of $\psi(\bm R)$
\begin{align}
    \psi_a(\bm R) &= \re [f (\bm R)] = \ln |\psi(\bm R)|\\
    \psi_\phi(\bm R) &= \im [f (\bm R)] = \arg \psi(\bm R).
\end{align}

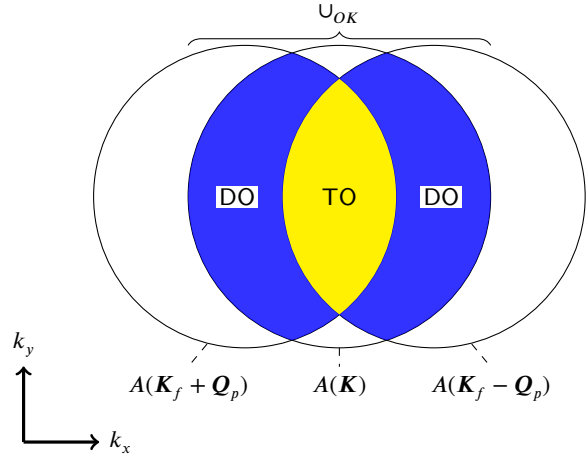
\begin{figure}
    \centering
        \begin{tikzpicture}
        \draw (0,0) circle (2);
        \draw[anchor=north, dashed] (0,-2) -- (0, -2.25) node {$A(\bm K)$};
        
        \draw (1.25,0) circle (2);
        \draw[anchor=north, dashed] (1.75, -1.95) -- (2, -2.25) node {$A({\bm K}_f - \bm Q_p)$};
        
        \draw (-1.25,0) circle (2);
        \draw[anchor=north, dashed] (-1.75, -1.95) -- (-2, -2.25) node {$A({\bm K}_f + \bm Q_p)$};
        
        \begin{scope}
        \clip (1.25,0) circle (2) (-1.25,0) circle (2);
        \fill[blue,opacity=0.8] (0,0) circle (2);
        
        \clip (1.25,0) circle (2);
        \clip (-1.25,0) circle (2);
        \fill[yellow,opacity=1] (0,0) circle (2);
        \end{scope}
        
        \draw (0,0) node {TO};
        \draw (1.35,0) node [fill=white, inner sep=1pt]{DO};
        \draw(-1.35,0) node [fill=white, inner sep=1pt]{DO};
        
        \begin{scope}[xshift=-5, yshift = -7]
        \draw[very thick, ->, anchor=south] (-4,-3) -- (-4 ,-2) node {$k_y$};
        \draw[very thick, ->, anchor=west] (-4,-3) -- (-3,-3) node {$k_x$};
        \end{scope}
        
        \draw [decorate,decoration={brace,amplitude=5pt}]
          (-2,2) -- (2,2) node[midway,yshift=12pt]{$\cup_{OK}$};
        \end{tikzpicture}
    \caption{The region $\uok$, a union of the double-overlap (DO) and triple-overlap (TO) regions in $(\kf, \qp)$-space. Weak object information only exists in this region of the four-dimensional space.}
    \label{fig:clark_zone}
\end{figure}

With this form $\psi(\bm R)$, we can simplify Eq. \ref{detector_intensity} following the same approach as Ref. \cite{rodenburg_experimental_1993}, but with an extra term included from our consideration of amplitude modulation by the sample. The Fourier transform of $\psi(\bm R)$ is approximately 
\begin{align}
    \Psi(\textbf{K}_f) &\approx \delta(\textbf{K}_f) + \Psi_a(\textbf{K}_f) + i\Psi_\phi(\textbf{K}_f), \\
    F(\kf) &= FT[f(\textbf{R}), \textbf{R}] =  \Psi_a(\textbf{K}_f) + i\Psi_\phi(\textbf{K}_f),
\end{align}
which gives four terms in Eq. \ref{detector_intensity} due to the product $\Psi(\textbf{K}_f - \textbf{K'}) \Psi^* (\textbf{K}_f - \textbf{K''})$. One term contains Dirac deltas in both integration variables, so both integrals can be immediately computed. One term is quadratic in $F(\textbf{K}_f)$ and is ignored. The remaining two contain just one Dirac delta, for example,
\begin{multline}
    \iint  A(\textbf{K'}) F(\textbf{K}_f - \textbf{K'}) A^* (\textbf{K''}) \delta (\textbf{K}_f - \textbf{K''}) \exp[2\pi i \textbf{R}_p \cdot (\textbf{K'} - \textbf{K''})] d\textbf{K'}d\textbf{K''} \\
    = \int  A(\textbf{K'}) A^* (\textbf{K}_f) F(\textbf{K}_f - \textbf{K'}) \exp[2\pi i \textbf{R}_p \cdot (\textbf{K'} - \kf)] d\textbf{K'}.
\end{multline}
Eliminating these last integrals motivates the first step of direct-interference ptychography post-processing: Fourier transforming with respect to the probe position $\textbf{R}_p$. The only appearance of $\textbf{R}_p$ is in the complex exponential, so taking a Fourier transform with respect to that variable produces a Dirac delta $\delta(\textbf{K'} - \textbf{K}_f + \textbf{Q}_p)$, where $\textbf{Q}_p$ is the reciprocal variable to $\textbf{R}_p$. This allows the last pair of integrals to be performed, resulting in
\begin{equation}
\begin{aligned}
        G(\textbf{K}_f, \textbf{Q}_p) &= FT[|M(\textbf{K}_f, \textbf{R}_p)|^2, \textbf{R}_p] \\
        &\approx A(\textbf{K}_f)|^2 \delta (\textbf{Q}_p) + A^* (\textbf{K}_f) A(\textbf{K}_f - \textbf{Q}_p) F (\textbf{Q}_p) + |A(\textbf{K}_f) A^*(\textbf{K}_f + \textbf{Q}_p) F^*(-\textbf{Q}_p)
\end{aligned}
\end{equation}
We can further simplify using Friedel's law, which states that $F(k)$, the Fourier transform of a real-valued function $f(x)$, has the property $F(k) = F^*(-k)$. Our $F(\qp)$ is composed of the Fourier transforms of the strictly-real-valued real and imaginary parts of $f(\bm R)$, so 
\begin{equation}
    F^*(-\qp) = \Psi_a^*(-\qp) - i\Psi_\phi^*(-\qp) = \Psi_a(\qp) - i\Psi_\phi(\qp) = F^*(\qp).
\end{equation}
This step is purely mathematical. We use no physical information about the sample, microscope, or sample-microscope-interaction physics. Then
\begin{equation} 
\begin{aligned}
    G(\textbf{K}_f, \textbf{Q}_p) = |A(\textbf{K}_f)|^2 \delta (\textbf{Q}_p) &+ A^* (\textbf{K}_f) A(\textbf{K}_f - \textbf{Q}_p) F (\textbf{Q}_p) \\
    &+ A(\textbf{K}_f) A^*(\textbf{K}_f + \textbf{Q}_p) F^*(\textbf{Q}_p) 
\end{aligned}
\end{equation}
or equivalently
\begin{equation}\label{fundamental_g}
\begin{aligned}
      G(\textbf{K}_f, \textbf{Q}_p) = |A(\textbf{K}_f)|^2 \delta (\textbf{Q}_p) &+  [A^* (\textbf{K}_f) A(\textbf{K}_f - \textbf{Q}_p) + A(\textbf{K}_f) A^*(\textbf{K}_f + \textbf{Q}_p)] \Psi_a(\textbf{Q}_p) \\
      &+ [A^* (\textbf{K}_f) A(\textbf{K}_f - \textbf{Q}_p) - A(\textbf{K}_f) A^*(\textbf{K}_f + \textbf{Q}_p)] i\Psi_\phi(\textbf{Q}_p).    
\end{aligned}
\end{equation}
This equation shows sample weak phase and amplitude information $\Psi_\phi(\qp)$ and $\Psi_a(\qp)$ are contained in regions of $(\kf,\qp)$-space defined by the magnitude of the aperture functions $A(\kf)$ and $A(\kf \pm \qp)$. An individual aperture magnitude $|A(\kf-\bm K_0)|$ is a circle of radius $\alpha/\lambda$ and center $\bm K_0$ in $\kf$ space. $|A(\kf)|$ is then a fixed disk about the $\kf$ axis in $(\kf,\qp)$-space, the central circle in Fig. \ref{fig:clark_zone}. The existence of sample information at a given spatial frequency $\qp$ depends on the overlap of the additional aperture functions $|A(\kf \pm \qp)|$ with the central disk. We define $\cup_{\text{OK}}$ as any $(\kf,\qp)$ where some overlap exists, or more formally, where $|A(\kf) A(\kf  - \qp)| + |A(\kf) A(\kf - \qp)| \not = 0$. This region is depicted for some $\qp$ such that $0 < |\qp| < \alpha/\lambda$ in Fig \ref{fig:clark_zone}. For $|\qp| \geq 2\alpha/\lambda$, the centers of the circles $|A(\kf \pm \qp)|$ are too far from the central disk for any overlap to occur, therefore there is no contrast observable from direct-interference ptychographic techniques above the $2\alpha$ limit. For $0 < |\qp| \leq 2\alpha/\lambda$, some of a single disk $|A(\kf \pm \qp)|$ overlaps with $|A(\kf)|$ where $|A(\kf \mp \qp)|$ does not. Such an area is called a double-overlap (DO) region. Once $0 \leq |\qp| \leq \alpha/\lambda$, the exterior disks can overlap each other inside the central disk, creating a triple-overlap (TO) region. This is the condition depicted in Fig. \ref{fig:clark_zone}.

Appropriately, $G$ in the WPOA can be recovered from Eq. \ref{fundamental_g} if the sample wavefunction does not affect the transmitted wavefront amplitude and $\Psi_a(\textbf{Q}_p) = 0$. Our derived factors of $\Psi_\phi(\qp)$ and $\Psi_a(\qp)$ in Eq. \ref{fundamental_g} are also proportional to Eqs. 13 and 19 of Ref. \cite{ma_information_2026}, specifically the phase and amplitude contrast transfer functions they derived in a more rigorous model of the STEM.

\section{Reconstruction in Theory} \label{recon_theory}

In the last section, we derived Eq. \ref{fundamental_g}, the distribution of weak sample information in 4D-STEM data. This does not, however, describe the relationship between the final reported reconstruction and the object wavefunction. For this, we need to model how direct-interference ptychography techniques analyze $G$ to extract sample information. In general, they follow an algorithm
\begin{enumerate}
    \item Move 4D-STEM data $|M(\kf, \rp)|^2$ to $(\kf,\qp)$-space via a Fourier transform with respect to $\rp$
    \item Identify an integration region $\mathcal{R}$
    \item Apply a filter $\eta(\kf, \qp)$
    \item Collapse the data to a 2D function $S(\qp)$ by integrating over all $\kf$ in $\mathcal{R}$
    \item Return to real space by inverse Fourier transforming $S(\qp)$ to produce the reconstruction $s(\rp)$
\end{enumerate}
DIP methods use different choices of integration regions $\mathcal{R}$ and filters $\eta$. This is the process outlined in Fig. \ref{fig:overview}. The result of the first step is $G(\kf,\qp)$. Steps 2-4 together produce the summation
\begin{equation} \label{general_sum}
\begin{aligned}
    S(\qp) = \int_\mathcal{R} |A(\kf)|^2 \delta(\qp) + \eta(\kf,\qp) [&\Gamma_+(\kf, \qp) \Psi_a(\qp) \\
    &+ i\Gamma_-(\kf, \qp) \Psi_\phi(\qp)]   d\kf
\end{aligned}
\end{equation}
where, for brevity, we write
\begin{align}
    &I_0 = \int_\mathcal{R} |A(\kf)|^2 d\kf \\
    &\Gamma_\pm(\kf,\qp) =  A^* (\textbf{K}_f) A(\textbf{K}_f - \textbf{Q}_p) \pm A(\textbf{K}_f) A^*(\textbf{K}_f + \textbf{Q}_p).
\end{align}
The resulting ptychographic reconstruction has the form
\begin{equation}
    s(\rp) = FT^{-1}[S(\qp), \rp].
\end{equation}

\subsection{Amplitude Reconstruction}

To perform one complex wavefunction reconstruction requires two separate direct-interference ptychographic reconstructions, one for the sample amplitude $s_a(\rp)$, and one for the sample phase $s_\phi(\rp)$. We begin with the amplitude reconstruction, where we choose $\mathcal{R} = \uok$ and $\eta = 1/2$. The Fourier-space sum for the amplitude reconstruction is
\begin{equation} \label{amp_S}
\begin{aligned}
    S_a(\qp) = I_0 \delta(\qp) &+ \frac{1}{2} \Psi_a(\qp) \int \Gamma_+(\kf,\qp) d\kf \\
    &+ \frac{i}{2} \Psi_\phi(\qp) \int \Gamma_-(\kf,\qp) d\kf  
\end{aligned}
\end{equation}
where we have dropped the subscript $\mathcal{R}$ because $\Gamma_\pm(\kf,\qp)=0$ for all pixels outside $\uok$ (all $\kf \not \in \mathcal{R}$). These integrals of $\Gamma_\pm$ over $\uok$ are equivalent to integrals over all $\kf$. The ptychographic amplitude reconstruction has the form
\begin{align}
    s_a(\rp) &= I_0 + FT^{-1} \left[\frac{1}{2} \Psi_a (\qp) \int \Gamma_+(\kf, \qp) d\kf \right] + FT^{-1} \left[\frac{i}{2} \Psi_\phi (\qp) \int \Gamma_-(\kf, \qp) d\kf \right] \\
    &= I_0 + \psi_a(\rp) \otimes_{\rp} FT^{-1}\left[ \int \Gamma_+(\kf, \qp) d\kf \right] + \frac{i}{2}\psi_\phi (\rp) \otimes_{\rp} FT^{-1} \left[ \int \Gamma_-(\kf, \qp) d\kf \right].
\end{align}
It is at this point we reap the benefit of including the triple-overlap region, which in and of itself is not a novel choice \cite{yang_enhanced_2016, bennemann_detective_2026, ooe_ultra-high_2021, ma_information_2026}. What has not yet been noted, to our knowledge, is that including the TO region allows considerable simplification of the function convolved with the sample information (the point spread function, see Section \ref{ctfs_and_ptfs}).
\begin{equation}
    \begin{aligned}
        \int FT^{-1}[\Gamma_+ (\kf,\qp), \rp] d\kf =& \int A^* (\textbf{K}_f) e^{-2\pi i \rp \cdot \kf} \int A(\textbf{Q'}) e^{2\pi i \rp \cdot \textbf{Q'}} d\textbf{Q'} d\kf \\
        &+ \int A(\textbf{K}_f) e^{2\pi i \rp \cdot \kf} \int A^*(\textbf{Q''})e^{-2\pi i \rp \cdot \textbf{Q''}} d\textbf{Q''} d\kf
    \end{aligned}
\end{equation}
\begin{equation}
    = a(\rp)\int A^* (\textbf{K}_f) e^{-2\pi i \rp \cdot \kf} d\kf + a^*(\rp)\int A(\textbf{K}_f) e^{2\pi i \rp \cdot \kf} d\kf
\end{equation}
\begin{equation} \label{invFT_of_gamma_plus}
    = a(\rp)a^*(\rp) + a^*(\rp)a(\rp).
\end{equation}
This integral producing twice $|a(\rp)|^2$ motivates the initial choice of $\eta = 1/2$. The result also implies
\begin{equation}
    \int FT^{-1}[\Gamma_- (\kf,\qp), \rp] d\kf = 0,
\end{equation}
which represents a much stronger analogue of the well-known loss of phase contrast in SSB inside the TO region when the microscope is in focus. This equation says \textit{no matter the aberrations}, $s_a(\rp)$ will not contain weak phase contrast. What remains is 
\begin{equation}
    s_a(\rp) = I_0 + |a(\rp)|^2 \otimes_{\rp} \psi_a(\rp).
\end{equation}
Recall $\psi_a(\rp)=\ln |\psi(\rp)|$, not the amplitude of $\psi(\rp)$. The natural log is the tradeoff for placing the amplitude inside the exponential so we could linearize it with the weak object approximation. Assuming small deviations from a unitary amplitude, we can expand $\ln x \approx x - 1$ about 1. Then
\begin{equation}
\begin{aligned}
    s_a(\rp) &= \ln(|\psi(\rp)|) \otimes_{\rp} |a(\rp)|^2 + I_0 \\
    &\approx |\psi(\rp)| \otimes_{\rp} |a(\rp)|^2 - 1 \otimes_{\rp} |a(\rp)|^2 + \int |A(\textbf{K}_f)|^2 d\kf 
\end{aligned} 
\end{equation}
and
\begin{equation}
    1 \otimes |a(\rp)|^2 = \int|a(\rp)|^2 d\rp.
\end{equation}
By Parseval's theorem, the two constant terms exactly cancel. We can then very succinctly describe the direct-interference ptychographic amplitude reconstruction as the convolution of the real-space STEM probe intensity with the amplitude of the sample wavefunction. 
\begin{equation} \label{amp_recon}
    s_a(\rp) = |\psi(\rp)| \otimes_{\rp} |a(\rp)|^2,
\end{equation}

\subsection{Phase Reconstruction}

We now turn to the phase reconstruction $s_\phi(\rp)$. Ideally, some careful choice of $\mathcal{R}$ and $\eta$ would cause the phase reconstruction to be independent of the sample amplitude regardless of probe aberrations. We were unable to analytically prove any particular combination of reconstruction parameters had this property. We employed the difference of the DO regions
\begin{equation} \label{phase_S}
    S_\phi(\qp) = \frac{1}{2}\int_{+ \qp \text{ DO}} G(\kf,\qp ) d\kf - \frac{1}{2}\int_{-\qp \text{ DO}} G(\kf,\qp ) d\kf,
\end{equation}
equivalent to applying a $\pi$ phase to one DO region in the method of Ref. \cite{pennycookEfficientPhaseContrast2015}. A broader version where the TO region is also negated and the classic SSB single-DO-region approach were equally successful at phase reconstruction (see Figs. \ref{fig:phase_recon_modes} and \ref{fig:recons}). We discuss the differences between these approaches and evidence for separability of phase and amplitude signal regardless of aberrations in Section \ref{discussion}. For the rest of this section, Eq. \ref{phase_S} is the basis for phase reconstruction, using the filter $\eta(\kf, \qp) = [1 - 2\Theta(|A(\kf + \qp)|)]/2$ and $\mathcal{R} = (+ \qp \text{ DO}) \cup (-\qp \text{ DO})$, that is, both DO regions. We proceed to derive a simplified form of $s_\phi(\rp)$, our ptychographic phase reconstruction.

In each DO region, only one term from Eq. \ref{fundamental_g} is nonzero 
\begin{equation}
\begin{aligned}
    S_\phi(\qp) = &\frac{1}{2}\int_{+ \qp \text{ DO}} A^*(\kf) A(\kf-\qp) F(\qp) d\kf \\
    &- \frac{1}{2}\int_{-\qp \text{ DO}} A(\kf) A^*(\kf+\qp) F^*(\qp) d\kf.
\end{aligned}
\end{equation}
To combine the integrals, we rewrite the bounds in terms of the whole $\uok$ and TO regions. Using the fact that the integrands are zero outside $\uok$,
\begin{equation}
\begin{aligned}
    S_\phi(\qp) = \frac{1}{2}F&(\qp) \int A^*(\kf) A(\kf-\qp) d\kf - \frac{1}{2}F(\qp) \int_\text{TO} A^*(\kf) A(\kf-\qp) d\kf \\
    - &\frac{1}{2}F^*(\qp) \int A(\kf) A^*(\kf+\qp) d\kf + \frac{1}{2}F^*(\qp) \int_\text{TO} A(\kf) A^*(\kf+\qp) d\kf.
\end{aligned}
\end{equation}
In terms of $\Psi_a(\qp)$ and $ \Psi_\phi(\qp)$,
\begin{align}
    S_\phi(\qp) &= \frac{1}{2}\Psi_a(\qp) \left[ \int \Gamma_- d\kf - \int_{\text{TO}} \Gamma_- d\kf \right] + \frac{i}{2} \Psi_\phi(\qp) \left[ \int \Gamma_+ d\kf - \int_{\text{TO}} \Gamma_+ d\kf \right] \\
    &=\frac{1}{2}\Psi_a(\qp) \int_{\text{DOs}} \Gamma_-(\kf,\qp) d\kf  + \frac{i}{2}\Psi_\phi(\qp) \int_{\text{DOs}} \Gamma_+ (\kf, \qp) d\kf.
\end{align}
This is a fairly intuitive result. Our amplitude and phase signals are weighted by integrals over our selected integration region.

Returning to real space,
\begin{equation}
    s_\phi (\rp) = \frac{1}{2}\psi_a(\rp) \otimes_{\rp} FT^{-1} \left[ \int \Gamma_- d\kf - \int_{\text{TO}} \Gamma_- d\kf \right] + \frac{i}{2}\psi_\phi(\rp) \otimes_{\rp} FT^{-1} \left[ \int \Gamma_+ d\kf - \int_{\text{TO}} \Gamma_+ d\kf \right].
\end{equation}
Using the identities $\int FT^{-1}[\Gamma_+] d\kf = |a(\rp)|^2$ and $\int FT^{-1}[\Gamma_-] d\kf = 0$, we find
\begin{multline} \label{phase_recon}
    s_\phi(\rp) = \frac{i}{2} \psi_\phi(\rp) \otimes_{\rp} |a(\rp)|^2  \\- \frac{i}{2}\psi_\phi(\rp) \otimes_{\rp} FT^{-1} \left[ \int_{\text{TO}} \Gamma_+ (\kf,\qp) d\kf\right] - \frac{1}{2}\psi_a(\rp) \otimes_{\rp} FT^{-1} \left[ \int_{\text{TO}} \Gamma_- (\kf, \qp) d\kf\right],
\end{multline}
or equivalently
\begin{equation}
    s_\phi(\rp) = \frac{i}{2}\psi_\phi(\rp)  \otimes_{\rp} FT^{-1} \left[ \int_{\text{DOs}} \Gamma_+ (\kf, \qp) d\kf \right] + \frac{1}{2}\psi_a(\rp) \otimes_{\rp} FT^{-1} \left[ \int_{\text{DOs}} \Gamma_-(\kf,\qp) d\kf \right].
\end{equation}
This is much more complicated than the amplitude reconstruction (Eq. \ref{amp_recon}), but still permits measurement of solely phase information. If ptychographic data is acquired in focus, or aberration corrected in postprocessing, 
\begin{equation}
    \Gamma_{\pm, \text{ a.c.}} = |A(\kf) A(\kf - \qp)| \pm |A(\kf) A(\kf + \qp)|.
\end{equation}
In the TO region, $\Gamma_{-, \text{ a.c.}} = 0$, eliminating any amplitude information in the phase reconstruction. In the DO regions, $\Gamma_{+, \text{ a.c.}} = 2$, so the final in-focus form of the phase reconstruction is
\begin{equation} \label{pure_phase_recon}
    s_{\phi, \text{ a.c}} (\rp) = i\psi_\phi(\rp)  \otimes_{\rp} FT^{-1} [C (\qp)] 
\end{equation}
where $C(\qp)$ is the area of the DO regions. As previously shown \cite{yangEfficientPhaseContrast2015}, this function can be written in terms of $\omega = |\qp| \lambda / \alpha$ as
\begin{equation}
    C(\qp) = C (\omega) = \frac{4}{\pi} \begin{cases}
    \cos^{-1} \left( \frac{\omega}{2}\right) - \cos^{-1} (\omega)  + \omega \sqrt{1-\omega^2} -\frac{\omega}{2} \sqrt{1-\left( \frac{\omega}{2}\right)^2}, 0\leq \omega \leq 1 \\
    \cos^{-1} \left( \frac{\omega}{2}\right) - \frac{\omega}{2} \sqrt{1-\left( \frac{\omega}{2}\right)^2}, 1 \leq \omega \leq 2
    \end{cases}.
\end{equation}

\subsection{Contrast Transfer and Point Spread Functions} \label{ctfs_and_ptfs}

Our derived amplitude and phase reconstructions (Eqs. \ref{amp_recon} and \ref{phase_recon}) are not perfect recreations of the object amplitude and phase. This is a well-known characteristic of linear imaging systems such as the STEM: the true object signal is obscured by its convolution with a point spread function (PSF) \cite{kirkland_advanced_2020,born_principles_2019}. The PSF is the image of a point source by the imaging system. We can also evaluate the system by its contrast transfer function (CTF), or the Fourier transform of the PSF. By the convolution theorem, the CTF multiplies the Fourier transform of the true signal in reciprocal space. The CTF is a key metric for judging the relative performance of various computational phase retrieval techniques \cite{varnavides_beyond_2026}. 

The separation of amplitude and phase information into real and imaginary parts by direct-interference ptychography requires distinct CTFs and PSFs for amplitude (ACTF, APSF) and phase retrieval (PCTF, PPSF). By definition, we can read off
\begin{align}
    \text{APSF}(\rp) &= |a(\rp)|^2 \\
    \text{PPSF}_{\text{a.c.}}(\rp) &= i FT^{-1}[C(\qp)] \\
    \text{ACTF}(\qp) &= \int \Gamma_+(\kf,\qp) d\kf \\
    \text{PCTF}_\text{a.c.}(\qp) &= iC(\qp).
\end{align}

\section{Discussion} \label{discussion}

Fig. \ref{fig:prelim_example} demonstrated the capacity for direct-interference ptychographic reconstructions of the wavefunction amplitude at ultra-low-dose. Now we demonstrate how direct-interference ptychography can outperform conventional imaging techniques at low and moderate doses. Reconstructions throughout the paper are performed with a minimally-adapted version of the publicly-available MATLAB code ptychoSTEM \cite{martinez_ptychostem_2022}. We use BF STEM, specifically integration of all intensity in the diffraction pattern within an angle $\alpha$ of the optical axis, to benchmark our amplitude reconstructions. A common mechanical step in DIP postprocessing is to crop data outside the bright-field disk, as no weak-phase-contrast electrons are scattered outside it \cite{rose_nonstandard_1976}. The conventional STEM technique we employ as a comparator should use the exact same electrons. BF STEM meets this requirement. Other STEM modalities which use different scattering geometries, such as dark-field STEM, can exhibit more interpretable amplitude contrast, but by probing different components of the specimen wavefunction amplitude.

We simulate a 4D-STEM measurement of 20-atomic-layer (7.8 nm) thick SrTi$\text{O}_3$ crystal oriented along the [100] axis in abTEM \cite{madsen_abtem_2021}. The probe is focused at the top surface with semi-convergence angles of $\alpha=$ 20, 25, and 30 mrad and electron doses of 1e2, 1e3, and 1e4 $\text{e}^-/\text{\AA}^2$. Full simulation parameters are listed in Appendix \ref{appendix}. Fig. \ref{fig:amps} shows the BF images and wavefunction amplitude reconstructions of the SrTi$\text{O}_3$ sample. DIP amplitude reconstructions consistently provide stronger contrast than BF STEM. Similar to the results in Fig. \ref{fig:prelim_example}, the background noise is much lower in DIP amplitude reconstructions, with the location of the heavier Sr atoms clearly identifiable down to doses of 1e2 $\text{e}^-/\text{\AA}^2$. This higher sensitivity allows moderate-dose reconstructions to detect the lighter Ti and O atoms, with distinct peaks visible between Sr atoms in the 20 mrad measurements at 1e3 and 1e4, and the 30 mrad measurement at 1e4 $\text{e}^-/\text{\AA}^2$.

\begin{figure}
	\centering
	\includegraphics[width=0.8\textwidth]{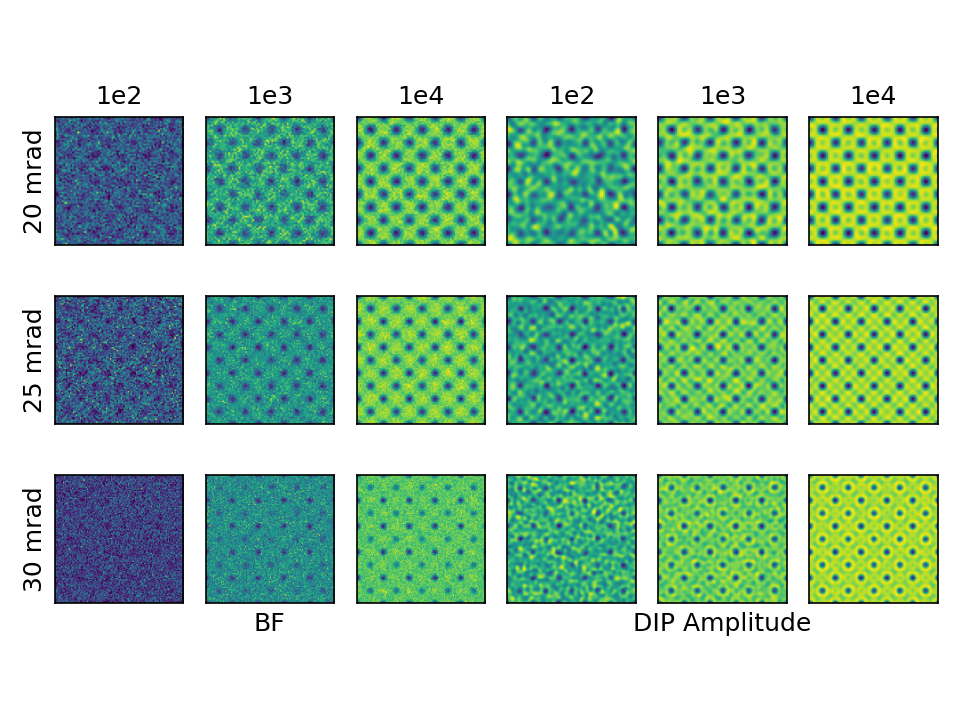}
	\caption{Simulated BF STEM measurements (leftmost three columns) and amplitude reconstructions by direct-interference ptychography (rightmost three columns) of SrTi$\text{O}_3$ along the [100] axis. Simulations are performed with varying finite electron dose and probe size. Each image is 19.5 by 19.5 \AA \; large.}
	\label{fig:amps}
\end{figure}

Our proposed DIP method and SSB ptychography produce similar phase reconstructions of the SrTi$\text{O}_3$ sample. SSB reconstructions are performed using the unaltered ptychoSTEM implementation. Fig. 5 presents the phase reconstructions for varying semi-convergence angles and electron doses. As expected, the lighter Ti and O atomic columns are far more discernible from the more strongly-scattering Sr atomic columns than in the amplitude reconstructions. At the lowest dose, it becomes difficult to differentiate between the different atomic species, although the lighter elements tend to appear smaller in the reconstructions than the Sr. The DIP and SSB phase reconstructions at 20 and 25 mrad are largely identical. SSB appears to have slightly better resolution at 30 mrad, with more localized atomic columns in the reconstructions. 

\begin{figure}
	\centering
	\includegraphics[width=0.8\textwidth]{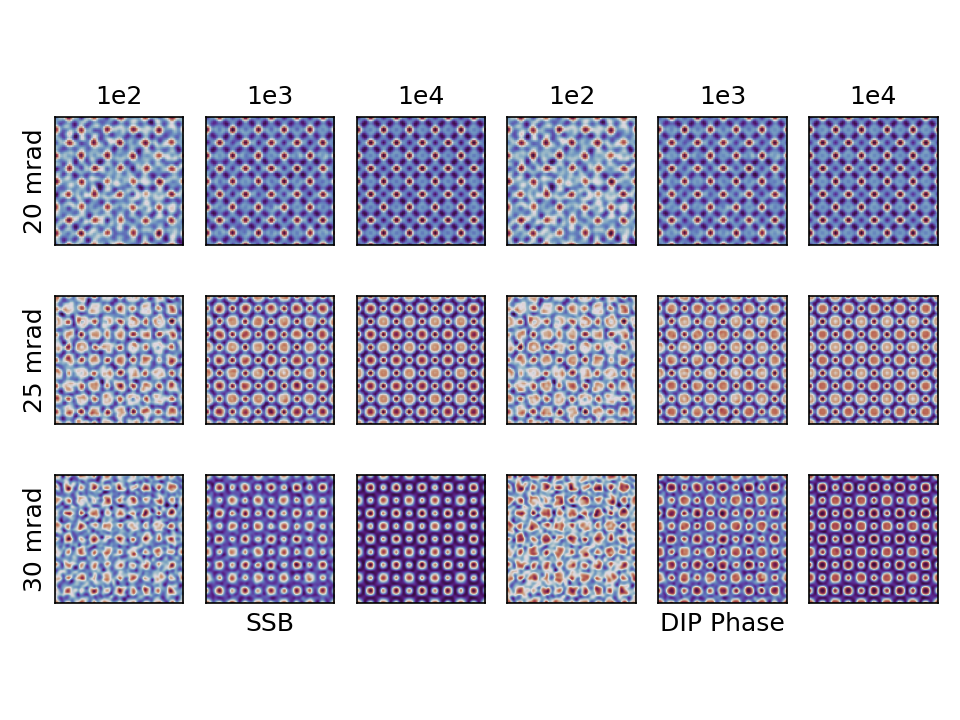}
	\caption{Phase contrast reconstructions of simulated SrTi$\text{O}_3$ along the [100] axis by single-sideband (leftmost three columns) and direct-interference ptychography (rightmost three columns). Simulations are performed with varying finite electron dose and probe size. Each image is 19.5 \AA \; by 19.5 \AA \; large.}
	\label{fig:phases}
\end{figure}

While the DIP and SSB phase reconstructions are very alike, our analysis predicts they should be identical. As the name suggests, the SSB ptychography extracts phase information from a single DO region. This is equivalent to using one of the terms in Eq. \ref{phase_S}. Without loss of generality, using the $+\qp$ DO region would produce an SSB reconstruction
\begin{equation}
    s_\text{SSB}(\rp) = [\psi_a(\rp) + i\psi_\phi (\rp)]\otimes_{\rp} FT^{-1} \left[ \int_{+\qp \text{ DO}} A^*(\kf) A(\kf-\qp) d\kf\right]
\end{equation}
under the WOA. In focus, 
\begin{equation} \label{ssb_recon}
    s_\text{SSB, a.c.}(\rp) = [\psi_a(\rp) + i\psi_\phi (\rp)]\otimes_{\rp} FT^{-1} \left[ C(\qp)\right].
\end{equation}
The imaginary part of the in-focus SSB reconstruction is identical to the imaginary part of our in-focus DIP reconstruction (see Eq. \ref{pure_phase_recon}).\footnote{Eq. \ref{ssb_recon} implies the sample wavefunction and amplitude can be simultaneously reconstructed by conventional SSB, an advantage over the two distinct amplitude and phase DIP reconstructions we propose. From simulation, we find the amplitude reconstruction proposed in Section \ref{recon_theory} to be more resilient to aberrations than the real part of the SSB reconstruction.} Yet the reconstructions differ. 

The discrepancy is due to a long-standing SSB convention. In the original SSB paper, Rodenburg, McCallum, and Nellist report the reconstruction amplitude and phase, although they never recommend a specific method of visualization \cite{rodenburg_experimental_1993}. The primary open-source SSB code ptychoSTEM, in both its MATLAB \cite{martinez_ptychostem_2022} and Python \cite{susi_open-source_2024} implementations, follows this course, computing the reconstruction amplitude and phase. Reviewing usage of the ptychoSTEM code in the literature \cite{hofer_phase_2024,li_improving_2025,yuan_atomically_2025} and our own previous work \cite{clark_effect_2023,dearg_stability_2025}, it seems that microscopists are predominantly reporting the SSB reconstruction phase
\begin{equation}
    \arg(s_\text{SSB}(\rp)) = \arctan \left( \frac{\psi_\phi (\rp)\otimes_{\rp} FT^{-1} \left[ C(\qp)\right]}{\psi_a(\rp)\otimes_{\rp} FT^{-1} \left[ C(\qp)\right]} \right).
\end{equation}
Interestingly, this implies that, despite always being rooted in the weak phase object approximation, SSB results have implicitly relied on the sample amplitude for contrast. If the WPOA was true, then $\psi_a(\bm \rp) = \ln |\psi(\rp)| = \ln 1 = 0$ and the phase of the reconstruction could only be $\pm \pi / 2$. 

\begin{figure}
    \centering
    \includegraphics[width=0.8\textwidth]{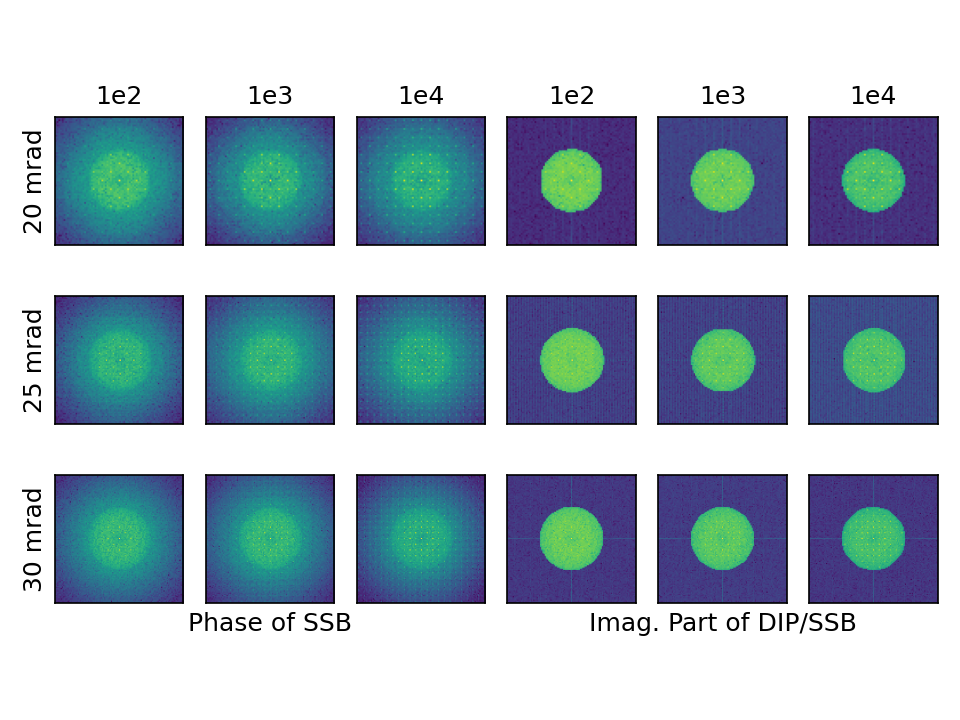}
    \caption{Fourier transforms of the phase contrast reconstructions of simulated SrTi$\text{O}_3$ in Fig. \ref{fig:phases} with varying finite electron dose and probe size. The leftmost three columns are transforms of reconstructions using the phase of the SSB reconstruction. The rightmost three columns are transforms of reconstructions using the imaginary part of our DIP reconstruction, which is identical to the imaginary part of the SSB reconstruction when both are acquired in focus.}
    \label{fig:phase_ffts}
\end{figure}

The apparent increased resolution of SSB relative our DIP method in Fig. \ref{fig:phases} is a consequence of plotting the phase of the SSB reconstruction. The ACTFs and PCTFs of the real and imaginary parts of the reconstruction are strictly limited to $0 \leq |\qp| \lambda \leq 2 \alpha$, but the SSB reconstruction phase has no such limitation. This can be observed in Fig. \ref{fig:phase_ffts}. Each image is the discrete Fourier transform of the corresponding reconstruction in Fig. \ref{fig:phases}. The Fourier transforms of the SSB reconstruction phase are much more diffuse whereas the DIP and SSB imaginary part reconstructions are strictly proscribed from spatial frequencies above the $2\alpha$ limit. 

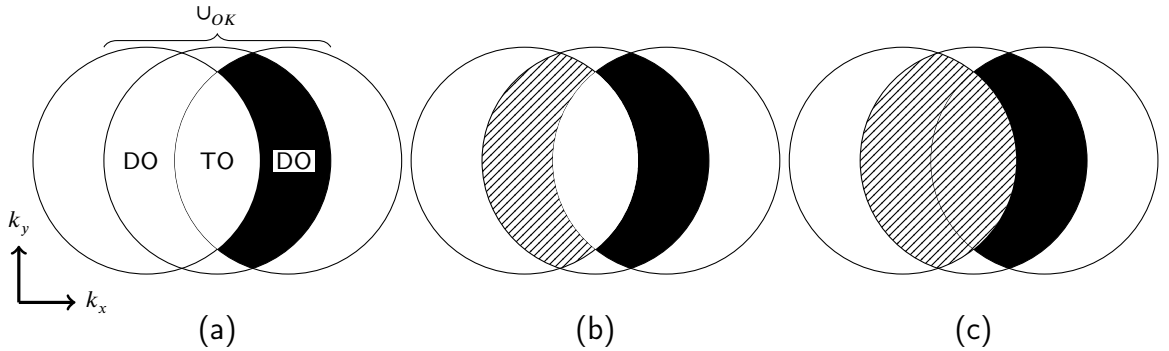
\begin{figure}
    \centering
       \begin{tikzpicture}
        \draw (-5,-2.25) node {\large (a)};
        \draw (0,-2.25) node {\large (b)};
        \draw (5,-2.25) node {\large (c)};
        \begin{scope}[xshift=-5cm, scale=0.75]
            \draw (0,0) circle (2);
            
            \draw (1.25,0) circle (2);
            
            \draw (-1.25,0) circle (2);
            \begin{scope}
            
            \clip (0,0) circle (2);
            \fill[black]  (1.25,0) circle (2);
            \clip (-1.25,0) circle (2);
            \clip (1.25,0) circle (2);
            \fill[white,opacity=1] (0,0) circle (2);
            \end{scope}
            
            \draw (0,0) node {TO};
            \draw (1.35,0) node [fill=white, inner sep=1pt]{DO};
            \draw(-1.35,0) node [fill=white, inner sep=1pt]{DO};
            
            \begin{scope}
            \draw[very thick, ->, anchor=south] (-3.5,-2.5) -- (-3.5,-1.5) node {$k_y$};
            \draw[very thick, ->, anchor=west] (-3.5,-2.5) -- (-2.5,-2.5) node {$k_x$};
            \end{scope}
            
            \draw [decorate,decoration={brace,amplitude=5pt}]
              (-2,2) -- (2,2) node[midway,yshift=12pt]{$\cup_{OK}$};
        \end{scope}
        \begin{scope}[xshift=0, scale=0.75]
            \draw (0,0) circle (2);
            
            \draw (1.25,0) circle (2);
            
            \draw (-1.25,0) circle (2);
            \begin{scope}
            
            \clip (0,0) circle (2);
            \fill[black]  (1.25,0) circle (2);
            \clip (1.25,0) circle (2);
            \clip (-1.25,0) circle (2);
            \fill[white,opacity=1] (0,0) circle (2);
            \end{scope}

            \begin{scope}
            
            \clip (0,0) circle (2);
            \fill[pattern=north east lines]  (-1.25,0) circle (2);
            \clip (-1.25,0) circle (2);
            \clip (1.25,0) circle (2);
            \fill[white,opacity=1] (0,0) circle (2);
            \end{scope}
        \end{scope}

        \begin{scope}[xshift=5cm, scale=0.75]
            \draw (0,0) circle (2);
            
            \draw (1.25,0) circle (2);
            
            \draw (-1.25,0) circle (2);
            \begin{scope}
            \clip (0,0) circle (2);
            \fill[black]  (1.25,0) circle (2);

            \clip (-1.25,0) circle (2);
            \clip (1.25,0) circle (2);
            \fill[white,opacity=1] (0,0) circle (2);
            \end{scope}

            \begin{scope}
            \clip (0,0) circle (2);
            \fill[pattern=north east lines]  (-1.25,0) circle (2);

            \end{scope}
        \end{scope}
    \end{tikzpicture}
    \caption{Phase reconstruction geometries producing apparently identical phase reconstructions (see Fig. \ref{fig:recons}). Reconstructions use data from solid-fill regions and negated data from diagonally-hatched regions.}
    \label{fig:phase_recon_modes}
\end{figure}

\begin{figure}
	\centering
	\includegraphics[width=0.8\textwidth]{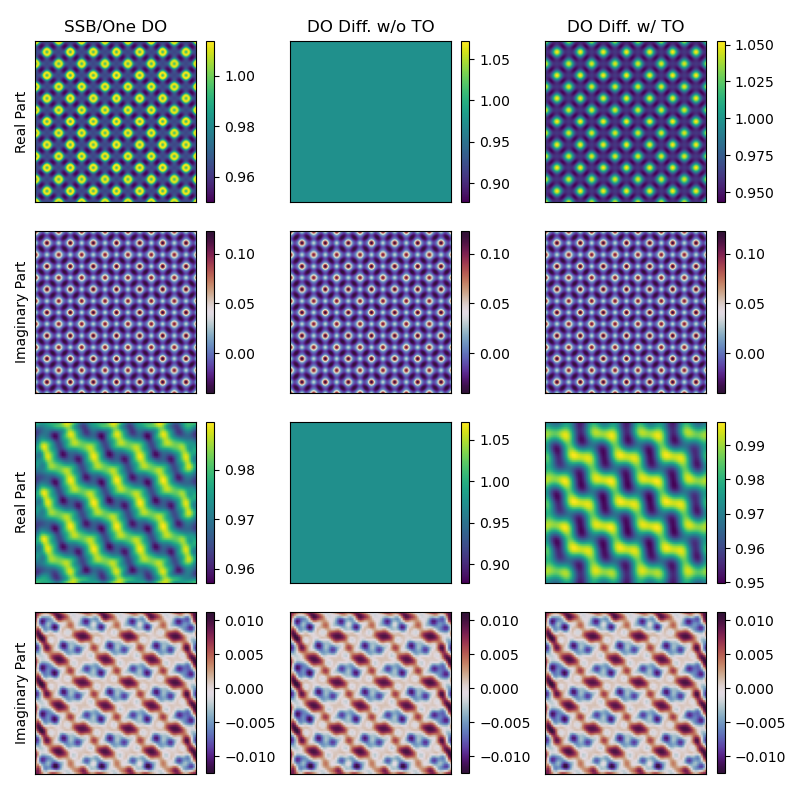}
	\caption{DIP reconstruction of simulated bilayer $\text{SrTiO}_3$ with varying selections of $\uok$ and two different probes. The columns correspond to reconstruction regions $\mathcal{R}$ depicted in Fig. \ref{fig:phase_recon_modes}. The first probe (top two rows) had 5 nm defocus. The second probe (lower two rows) was highly asymmetric via imposed aberrations of $C_{21} = 5$ nm, $\phi_{21} = \pi / 7$, and a beam tilt of 0.1 mrad.}
	\label{fig:recons}
\end{figure}

While we are well-motivated to use the imaginary part instead of the phase of an SSB reconstruction, there is not overwhelming theoretical evidence as to why we should use both DO regions to isolate phase information. Three geometries $\mathcal{R}$ appear identically capable of separably reconstructing the sample phase regardless of probe aberrations. They are depicted in Fig. \ref{fig:phase_recon_modes}: (a) the single sideband or one DO region, (b) the difference of the DOs, and (c) the difference of one DO and the TO regions from the other DO region. We have been unable to analytically prove separability independent of aberrations despite the indicative results in Fig. \ref{fig:recons}. Therein, phase reconstructions of simulated bilayer SrTi$\text{O}_3$ (simulation parameters detailed in Appendix \ref{appendix}) are performed on data acquired with an aberrated probe. Any aberrations breaks the assumption used to show separable phase retrieval in Section \ref{recon_theory}. The first simulated probe was 5 nm defocused. The real and imaginary parts of the three reconstruction approaches are depicted in the first and second rows of Fig. \ref{fig:recons} respectively. Taking the difference of DO regions (middle column) produces no contrast in the real part, which could signify the elimination of amplitude information regardless of aberrations. However, the imaginary parts of all three methods are indistinguishable to the eye and differ per pixel by less than $\pm5$e-9~rad. This implies amplitude information is confined to the real part of these reconstructions. We next reconstructed data acquired with an asymmetric probe to investigate the effects of breaking the probe symmetry about the optical axis. Despite imposing a beam tilt of 0.1 mrad and axial coma via parameters $C_{21} = 5$ nm and $\phi_{21}= \pi / 7$ rad, the same behavior as the symmetric probe case is observed across the reconstruction methods in the third and fourth rows of Fig.~\ref{fig:recons}. The imaginary parts of the reconstructions remain functionally equivalent. This is very promising for demonstrating DIP can reconstruct the unmixed sample phase and amplitude information regardless of the probe aberrations.

\section{Conclusion}

Amplitude contrast, unlike its counterpart phase contrast, is not difficult to measure in STEM. To fully understand our samples, however, we must be capable of disentangling the amplitude from phase signal. In this manuscript, we provide a step in this direction via non-iterative direct-interference ptychography, demonstrating the separable measurement of weak phase and amplitude contrast. Our amplitude reconstructions provide increased signal-to-noise over conventional amplitude measurement methods. This progress can be immediately applied to other direct-interference ptychographic techniques to better represent the physical structure underlying our reconstructions. OBF is an exciting candidate for integrating the WOA, as, uniquely among DIP techniques, its filter $\eta$ is designed to correct for the effective defocus of the beam along the optical axis and the modulation of contrast transfer due to Poisson noise. 

More work is needed to understand the theoretical performance of our approach in experiments with Poisson noise. We plan to investigate the spectral signal-to-noise ratio \cite{varnavides_beyond_2026} or detective quantum efficiency \cite{bennemann_detective_2026} of this technique in future work. As the DQE of SSB is half of its CTF \cite{bennemann_detective_2026}, we do not expect noise to fundamentally alter our conclusions. The impacts of partial spatial and temporal coherence should also be studied.


\appendix
\section{Appendix}
\label{appendix}

Simulations of DIP measurements of SrTi$\text{O}_3$ were performed in abTEM. The sample was oriented along the [100] axis and was five unit cells by five unit cells across. The potential had a sampling of 0.05 \AA. An 80 keV probe was focused at the top sample surface for $\alpha = 20$, 25, and 30 mrad. For all semi-convergence angles, the probe step size was half the Nyquist frequency. This was 0.26 \AA \; at 20 mrad, 0.21 \AA \; at 25 mrad, and 0.17 \AA \; at 30 mrad. The detector max angle was 35 mrad and uniform resampling was used.

The probe was in focus on a simulated sample 20 atomic layers thick for Figs. \ref{fig:amps} and \ref{fig:phases}. The simulated sample for Fig. \ref{fig:recons} was two atomic layers thick. For the first simulation on that sample, the defocus was set to 5 nm ($C_{10} = -50$ \AA). For the second simulation, there was no defocus, a beam tilt of 0.1 mrad, and aberration parameters $C_{21} = 5$ nm and $\phi_{21} = \pi / 7$ rad.

\printcredits

\bibliographystyle{unsrtnat}

\bibliography{references.bib}

\end{document}